\documentstyle[aps,prl,epsf]{revtex}
\begin{document}
\twocolumn[\hsize\textwidth\columnwidth
\hsize\csname@twocolumnfalse\endcsname
\draft 
\title{Answer to the comment of Chudnovsky: On the square-root time relaxation in 
molecular nanomagnets}
\date{\today}
\maketitle
%\begin{abstract}
%\end{abstract}
%\bigskip
%\pacs{PACS numbers: 75.45.+j, 75.60Ej}
\vskip1pc]
\narrowtext

In Ref. \cite{PRL_hole_dig},we presented a new technique, which we call the
{\it hole digging} method that can be used to observe the time evolution of
molecular states in crystals of molecular clusters.
It allows us to measure the statistical distribution of magnetic bias
fields that arise from the weak dipole fields of the clusters themselves. A
hole can be 'dug' into the distribution by depleting the available spins at
a given applied field.
Our method is based on the simple idea that after a rapid field change, the
resulting short time relaxation of the magnetization is directly related to the
number of molecules which are in resonance at the given applied field.

Prokof'ev and Stamp have suggested that the short time relaxation should
follow a $\sqrt{t}-$relaxation law \cite{Prok}. 
However, the hole digging method should
work with any short time relaxation
law \cite{remark1}. For the molecular cluster system Fe$_8$ it is an
experimental fact that the short time relaxation follows to a good 
approximation the $\sqrt{t}-$relaxation law, regardless whether we start from
the saturation magnetization \cite{Ohm}, or from an annealed state
\cite{PRL_hole_dig,remark2}. This is true as long as the measurements are made
below 400 mK, and therefore the relaxation is purely due to quantum
tunneling, and not to thermal activation \cite{Sangregorio97}.

An important result of our paper was the observation that the hole line
width becomes independent of the initial value of the magnetization for
small values of the later. We suggested in Ref. \cite{PRL_hole_dig} that
this {\it intrinsic} hole line width is directly related to the
inhomogeneous level broadening due to nuclear spins as predicted by
Prokof'ev and Stamp. Since the publication of our article, we have made new
measurements on isotopically substituted samples of Fe$_8$. Samples enriched
with $^{57}$Fe had a larger hole line width, and samples where H is replaced with
deuterium had a more narrow hole line width \cite{PRL2000}.
These measurements confirm our hypotheses, and are in quantitative
agreement which  numerical simulations which takes into account the altered
hyperfine coupling \cite{PRL2000}.

For a \emph{saturated} sample, the Prokof'ev--Stamp theory allows us to estimate 
the tunneling matrix element $\Delta$. Using Eqs. (3), (9) and (12) of \cite{Prok},
and integration, we find $\int \Gamma_{\rm sqrt}d\xi = 
c \frac{\xi_0}{E_D} \frac{\Delta^2}{\hbar}$, where $c$ is a constant of the order 
of unity which depends on the sample shape. With $E_D$ = 15 mT, 
$\xi_0$ = 0.8 mT, $c$ = 1 and $\Gamma_{\rm sqrt}$ 
from Fig. 3 in \cite{PRL_hole_dig}, 
we find $\Delta = 0.6 \times 10^{-7}$ K 
which is close to the result of $\Delta = 0.5 \times 10^{-7}$ K obtained 
by using a Landau Zener method \cite{Science}.

As for Mn$_{12}$ in  Ref \cite{EPL_Mn12} we did \emph{not} claim that the
relaxation follows the $\sqrt{t}-$ relaxation law. It is well known 
that the situation in this sample is
more complicated due to the fact that there are several coexisting species of
Mn$_{12}$ in any crystal, each with different relaxation times.
In Ref \cite{EPL_Mn12} we were able to isolate one faster relaxing species and we
stated in that case, that the relaxation can be {\it approximately} fit to
the $\sqrt{t}-$ relaxation law, but in fact is better fit to a power
law $t^\alpha$ with $0.3 < \alpha < 0.5$ (depending on the applied field). 
We applied the
hole digging method to this species, and found evidence for
intrinsic line broadening below 0.3K which we suggest comes from nuclear
spins in analogy with Fe$_8$. It should be mentioned that a constant 
external field can only shift the internal fields.
We also measured the relaxation of Mn$_{12}$ 
at higher temperature (0.04 - 5 K) and found {\it no evidence what so ever}
for a short time $\sqrt{t}-$ relaxation. 

Finally, we emphasize that the measurements of short time 
relaxation allows us to study the
time evolution of a well-defined initial state, 
whereas the interpretation of 
long time relaxation data are far more difficult due to the development of complex intermolecular correlations during the relaxation process, which are not yet well understood \cite{Prok,Ohm}.
\\
\\
W. Wernsdorfer$^1$, C. Paulsen$^2$, R. Sessoli$^3$\\
\\
$^1$Lab. Louis N\'eel and $^2$CRTBT\\
CNRS, BP 166, 38042 Grenoble Cedex 9, France\\
$^3$Dept. of Chemistry, University of Firenze, \\
via Maragliano 72, 50144 Firenze, Italy


\begin{references}
\bibitem{remark1} For example a power law $M \sim (\Gamma_{short}
t)^\alpha$, where $\Gamma_{short}$ is a characteristic short time
relaxation rate.
\bibitem{remark2} For thermal annealed samples, we found experimentally that the smaller the difference $(M_{eq} - M_{in})$, the better is the $\sqrt{t}-$fit.
\bibitem{Science} W.Wernsdorfer and R. Sessoli, Science {\bf 284}, 133 (1999). 
In this paper, $\Delta$ is the tunnel spitting which is two times the tunneling 
matrix element which was used in \cite{PRL_hole_dig}.
\end{references}
\end{document}